\documentclass[fleqn,10pt]{wlscirep}

\title{Structural Pathways Supporting Swift Acquisition of New Visuo-Motor Skills}

\author[1,2,3]{Ari E. Kahn}
\author[4]{Marcelo G. Mattar}
\author[3,5,2]{Jean M. Vettel}
\author[6]{Nicholas F. Wymbs}
\author[5]{Scott T. Grafton}
\author[2,7,8]{Danielle S. Bassett}
\affil[1]{Department of Neuroscience, University of Pennsylvania, Philadelphia, PA 19104 USA}
\affil[2]{Department of Bioengineering, University of Pennsylvania, Philadelphia, PA 19104 USA}
\affil[3]{Human Research and Engineering Directorate, U.S. Army Research Laboratory, Aberdeen, MD 21001 USA}
\affil[4]{Department of Psychology, University of Pennsylvania, Philadelphia, PA 19104 USA}
\affil[6]{Department of Physical Medicine and Rehabilitation, Johns Hopkins University, Baltimore, MD 21218 USA}
\affil[5]{Department of Psychological and Brain Sciences, University of California, Santa Barbara, CA 93106 USA}
\affil[7]{Department of Electrical and Systems Engineering, University of Pennsylvania, Philadelphia, PA 19104 USA}
\affil[8]{To whom correspondence should be addressed: dsb@seas.upenn.edu.}

\newpage
\begin{abstract}
Human skill learning requires fine-scale coordination of distributed networks of brain regions linked by white matter tracts to allow for effective information transmission. Yet how individual differences in these anatomical pathways may impact individual differences in learning remains far from understood. Here, we test the hypothesis that individual differences in structural organization of networks supporting task performance predict individual differences in the rate at which humans learn a visuo-motor skill.  Over the course of 6 weeks, twenty healthy adult subjects practiced a discrete sequence production task, learning a sequence of finger movements based on discrete visual cues. We collected structural imaging data, and using deterministic tractography generated structural networks for each participant to identify streamlines connecting cortical and sub-cortical brain regions. We observed that increased white matter connectivity linking early visual regions was associated with a faster learning rate. Moreover, the strength of multi-edge paths between motor and visual modules was also correlated with learning rate, supporting the potential role of extended sets of polysynaptic connections in successful skill acquisition. Our results demonstrate that estimates of anatomical connectivity from white matter microstructure can be used to predict future individual differences in the capacity to learn a new motor-visual skill, and that these predictions are supported both by direct connectivity in visual cortex and indirect connectivity between visual cortex and motor cortex.
\end{abstract}

\begin{document}

\flushbottom
\maketitle
%
%

\thispagestyle{empty}

\section*{Keywords:} diffusion MRI, motor sequence learning, graph theory, polysynaptic networks, discrete sequence production


\newpage
\section*{Introduction}

Human skill learning is a complex phenomenon that involves the fine-scale coordination of disparate cortical and subcortical regions \citep{Dayan:2011cz}. This coordination critically depends on the effective transmission of information across white matter tracts, which link distant brain regions in cortico-cortical networks and cortico-subcortical loops \citep{Lynch2006}. Lesions or injuries to these interconnected tracts -- particularly in motor and visual systems -- can directly cause deficits in skill learning \citep{Ding2001}. The exact extent of these deficits is difficult to predict, largely due to the fact that white matter tracts form a complex interconnected network \citep{Sporns2005}. Damage to this network can have broadly distributed repercussions on processing, causing loss of information transmission \citep{Scantlebury2014}, or detrimental alterations in transmission patterns \citep{Crofts2011}.

The interconnected nature of white matter tracts not only complicates response to injury, but it also forms a fundamental substrate for individual differences in brain anatomy that may have non-trivial effects on cognition and behavior. White matter connectivity displays large-scale differences across individuals \citep{Bassett2011structure}, being modulated by age \citep{Betzel2014}, gender \citep{Tunc2016,Ingalhalikar2014}, genetics \citep{Hong2015}, and prior experience \citep{Sampaio2013,Scholz2009}. How these individual differences may account for individual differences in skill learning is not fully understood.  Gaining such an understanding could directly inform therapeutic interventions to enhance recovery of motor skills after brain injury \citep{Tomassini2011}, and furthermore could potentially inform training paradigms to enhance motor-visual expertise in healthy individuals \citep{Neumann2016}.

\begin{figure}[!ht]
\centering
\includegraphics{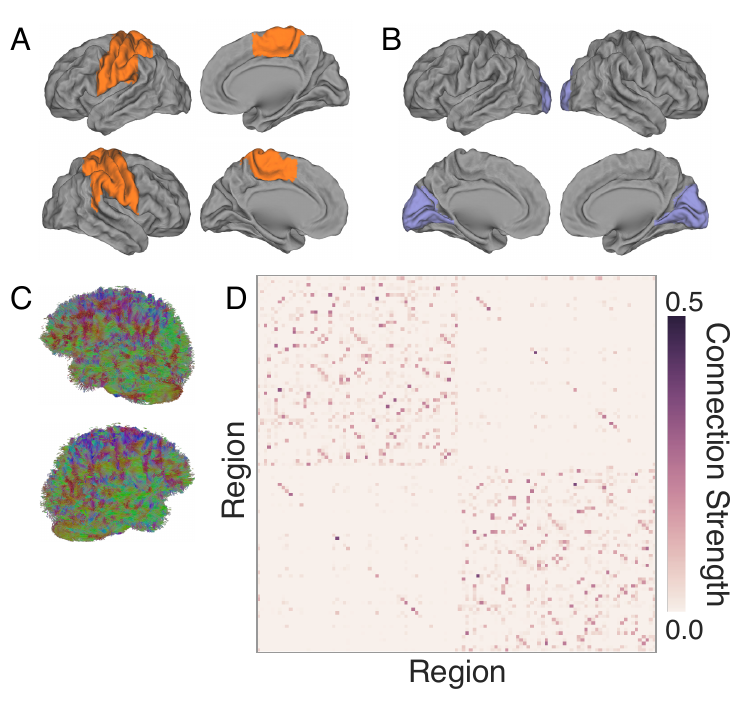}
\caption{\textbf{Structural Connectivity in Motor and Visual Networks of Interest.} \emph{(A--B)} Previous research suggests that increased skill on the discrete sequence production (DSP) task requires concerted functional network changes in distributed regions of motor \emph{(A)} and visual \emph{(B)} systems \citep{Bassett2015,Bassett:2013hs}; see Table~\ref{Tab1} for region names. \emph{(C)} To assess structural correlates of individual differences in learning rate on the DSP task, we performed deterministic diffusion imaging tractography on 4 scans dispersed evenly throughout the 6 weeks of training. \emph{(D)} We constructed structural networks using diffusion imaging tractography and the 111 cortical and subcortical regions in the Harvard-Oxford atlas to examine individual variability in connectivity strength. We also show that our results are robust across atlases, replicating our findings in the 90 cortical and subcortical region parcellation of the automated anatomical labeling (AAL) atlas.
\label{mv_dti}}
\end{figure}

Here, we examine if connectivity networks defined by diffusion MRI are predictive of individual differences in the rate at which subjects acquire a simple visuo-motor task \citep{Wymbs:2015ds}. In a discrete sequence production (DSP) task, subjects perform a sequence of finger movements based on visual cues \citep{Rhodes2004}. Once the correct key for each movement is pressed, the visual cue for the next sequence element is presented without delay. Consequently, a DSP task allows subjects to develop exceptionally fast, contiguous movements, much like an expert pianist performing a keyboard arpeggio. Efficient acquisition of this specific visuo-motor skill requires a gradual autonomy of visual and motor functional subnetworks \citep{Bassett2015,Bassett:2013hs} (Fig.~\ref{mv_dti}A--B). Initially, a person relies on the visual cue to perform a finger movement, an action that requires integration between motor and visual cortices; however, once a sequence becomes overlearned, a subject has mastered direct motor-motor associations where a given finger movement is the cue for the next finger movement.

These functional network changes may depend on underlying structure, shown to be a fundamental driver of brain dynamics at rest \citep{Becker2015,Goni2014,Honey2009} and during task performance \citep{Hermundstad2013,Hermundstad2014,Jarbo:2015bb,Smith:2009bh,Osher:2016cv}. Furthermore, individual variability in behavior has been linked to differences in structural networks \citep{JohansenBerg:2010ik}, and IQ and motor speed have been associated with greater white matter connectivity \citep{Li2009} and fractional anisotropy (FA) \citep{Hirsiger2016}. Prior work in word learning tasks also suggests that increased myelination, axonal diameter, and fractional anisotropy in tracts implicated in task processing are associated with better performance \citep{Barroso2013,Wong2011}. Building on these prior studies, we hypothesized that individuals with greater structural connectivity in motor and visual cortices (and particularly in primary motor and visual cortices) would show faster learning rates than individuals with less connectivity. We also set out to test whether these structural differences remained constant over the 6 weeks of practice \citep{LeBihan:2012cy} or changed appreciably with training \citep{Scholz2009,blumenfeld2011diffusion,taubert2012learning}. Finally, due to the prevalence of physically extended sets of polysynaptic connections in the visual-motor system, we hypothesized that individual differences in long distance walks on the graph of structural connections between visual and motor cortex would correspond to individual differences in learning rate.

To address these hypotheses, we examined diffusion tensor imaging data acquired from 20 healthy young adult subjects over the course of 6 weeks of training on the DSP task \citep{Bassett:2013hs,Bassett2014,Bassett2015,Wymbs:2015ds}. Subjects were scanned in 4 separate sessions, including a scan on day 1 before training began and then a scan approximately every 2 weeks. Between scanning sessions, subjects practiced a set of ten-element sequences at home using a program installed on their laptop computers, and behavioral performance was assessed by calculating the movement time (MT) for each sequence defined as the duration between the first button press and the last button press in the sequence. The learning rate for each participant was computed as the first exponential drop-off parameter in a double-exponential fit of the MT as a function of trials practiced across the entire 6 weeks of training. To compare individual differences in learning rate to the organization of white matter connectivity, we generated structural networks from the 4 diffusion tensor imaging scans using a deterministic fiber tracking algorithm (Fig.~\ref{mv_dti}C), which provided estimates of the number of streamlines connecting pairs of cortical and subcortical regions derived from brain atlases (Fig.~\ref{mv_dti}D). We observe three main results: individual differences in learning rate are significantly correlated with white matter connectivity in visual (but not motor) cortex, these relationships are consistent across the 6 weeks of task practice, and individuals with faster learning rates also show greater \textit{walk strength} linking motor and visual cortices, a measure suggesting increased strength of polysynaptic pathways.

\textbf{Figure 1}

\section*{Materials and Methods}

\subsection*{Participants and Experimental Design:}

\textit{Participants:} Twenty-two right-handed participants (13 females and 9 males; mean age, 24 years) volunteered and provided informed consent in writing in accordance with the guidelines of the Institutional Review Board of the University of California, Santa Barbara. All had normal or corrected vision and no history of neurological disease or psychiatric disorders. We excluded two participants because one participant failed to complete the experiment and the other had excessive head motion (persistent head motion greater than 5mm during the MRI scanning). We also had technical problems for two participants and were unable to collect DTI data during the pretraining session for Scan 1. Finally, for one additional subject, Scan 1 was removed due to the total of estimated streamlines differing by more than 3 standard deviations from the subject mean. Therefore, the structural analysis includes 17 participants for Scan 1 and 20 participants for Scans 2--4.

\begin{figure}[!ht]
\centering
\includegraphics{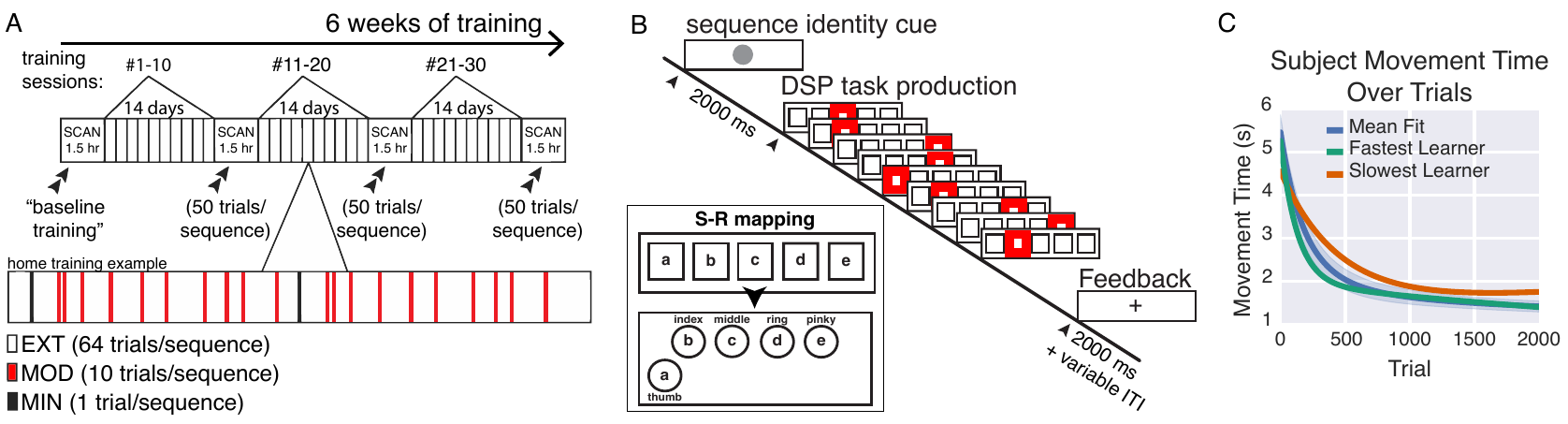}
\caption{\textbf{Overview of training, task paradigm, and MT estimation.} \emph{(A)} Training Schedule. Subjects underwent four scans, each approximately two weeks apart. Subjects practiced once a day for at least 10 days between each scanning session. \emph{(B)} Subjects viewed a screen on which stimuli were displayed. Each sequence was preceded with the display of a sequence identity cue, which informed the subject which of six sequences would follow. During the sequence, subjects saw five horizontally arranged squares. For each element of the sequence, one box was highlighted for the subject, providing information on the key to press. Upon completion of the task, a fixation cross was displayed for a short Inter-Trial Interval (ITI), and every 10 trials performance feedback was provided. The squares were spatially mapped onto a key-pad, one corresponding to each finger in addition to the thumb (see insert). \emph{(C)} Double exponential fit of MT to the number of trials practiced. The fit is shown for the fastest learner, the slowest learner, and the mean across all subjects.
\label{training_task_fig}}
\end{figure}

\textit{Experimental setup and procedure:} The DSP training protocol occurred over a six week period with four MRI scanning sessions spaced two weeks apart on Day 1, Day 14, Day 28, and Day 42 (Fig.~\ref{training_task_fig}A). On Day 1 of the experimental protocol, the participants completed their first MRI session, Scan 1, and the experimenter (N.F.W.) installed the training module on the participant's personal laptop and taught them how to use it for at-home training sessions. Participants were required to do the training for a minimum of 10 out of the 14 days in each 2-week period between the subsequent scanning sessions for Scans 2-4. All participants completed the full expected training regimen; none completed less than 10 full training sessions.

In their at-home training sessions, participants practiced a set of 10-element sequences using their right hand in a discrete sequence production (DSP) task \citep{Mattar2015,Bassett2015,Bassett:2013hs,Bassett2014,Wymbs:2015ds}. Sequences were presented using a horizontal array of five square stimuli, and the key responses were mapped from left to right, such that the thumb corresponded to the leftmost stimulus and the pinky finger corresponded to the rightmost stimulus (Fig.~\ref{training_task_fig}B). A square highlighted in red served as the imperative stimulus, and the next square in the sequence was highlighted immediately after each correct key press. If an incorrect key was pressed, the sequence was paused at the error and restarted upon the appropriate key press. Participants had an unlimited amount of time to respond and complete each trial.

\textbf{Figure 2}

Each practice trial began with the presentation of a sequence-identity cue that identified one of six sequences. These six sequences were presented with three different levels of exposure, in order to acquire data over a larger range of learning stages while controlling for the effect of scanning day (Table~\ref{t:training}). The two extensively trained (EXT) sequences were identified with a colored circle (cyan for sequence A and magenta for B), and they were each practiced for 64 trials during every at-home training session. The two moderately trained (MOD) sequences were identified by triangles (red for sequence C and green for D) and each practiced for 10 trials in every session. The two minimally trained (MIN) sequences were identified by black outlined stars (filled with orange for sequence E and white for F) and only practiced for 1 trial each during the at-home training sessions. Participants were given feedback every ten trials that reported the number of error-free sequences and the mean time required to complete them.

During each of the four MRI scanning sessions, we collected functional Echo Planar Imaging EPI data and structural imaging data from MPRAGE and DTI scans. In the functional runs, participants performed 300 trials of the self-paced DSP task using the same block structure with feedback as the at-home practice sessions, but the sequences were presented equally for a total of 50 trials for each of the six trained sequences. We have previously reported results from functional analyses \citep{Mattar2015,Bassett2015,Bassett:2013hs,Bassett2014,Wymbs:2015ds}. In this paper, we analyze the structural data and examine individual variability in structural connections among the distributed motor and visual regions of interest that were derived directly from the functional neuroimaging studies of this same dataset \citep{Bassett2015}. In this previous work, a set of motor and visual regions that formed functional modules were identified in a data-driven fashion whose task-based modulation tracked the effects of training. Here we build on the identification of these regions of interest by studying their structural connectivity derived from diffusion imaging.

\textbf{Table 1}

\subsubsection*{Estimating Learning Rates for Individual Participants}

For each sequence, we defined the MT as the duration between the time of the first button press and the time of the last button press. For the set of sequences of a single type (i.e., sequence A,B,C,D,E, and F), we estimated the learning rate by fitting a double exponential function to the MT data \citep{Schmidt:2005uh},\citep{Rosenbaum:2014ta} using a robust outlier correction in MATLAB (using the function ``fit.m'' in the Curve Fitting Toolbox with option ``Robust'' and type ``Lar''): $\text{MT} = D_1e^{-t\kappa} + D_2e^{-t\lambda}$ , where $t$ is time, $\kappa$ is the exponential drop-off parameter (which we called the \emph{learning rate}) used to describe the fast rate of improvement, $\lambda$ is the exponential drop-off parameter used to describe the slow, sustained rate of improvement, and $D_1$ and $D_2$ are real and positive constants. The magnitude of $\kappa$ indicates the steepness of the learning slope: individuals with larger $\kappa$ values have a steeper drop-off in MT, suggesting that they are faster learners \citep{Dayan:2011cz,Yarrow:2009jn}. The decrease in MT has been used to quantify learning for several decades \citep{Snoddy:1926gl,CROSSMAN:2010bw}. Several functional forms have been suggested for the fit of MT \citep{Newell:1993uk,Heathcote:2000uy}, and variants of an exponential are viewed as the most statistically robust choices \citep{Heathcote:2000uy}. In addition, the fitting approach that we used has the advantage of estimating the rate of learning independent of initial performance or performance ceiling. For the purpose of measuring effects on learning rate, we used the average value of $\kappa$ for the two extensively-trained sequences, for which we had the greatest number of trials practiced (Table \ref{t:training}).

While we do not have explicit information on the computing power of each subject's laptop, the learning rate that we study is independent of the starting MT, the ending MT, and the mean MT. Instead, it is a measure of the rate of change in MT. Thus, any differences in computing power cannot be used to explain the results. Moreover, we should mention that error rates on this task are on the order of $1 \times 10^{-3}$ \citep{Bassett2015}, and error rates are not significantly correlated with learning rates ($r=0.34$, $p=0.13$) \citep{Bassett2015}.

\subsection*{Neuroanatomical Data and Associated Methods}

In this section, we briefly describe the neuroanatomical data acquired from participants, as well as computational methods associated with data preprocessing, structural network construction, and statistical analyses.

\subsubsection*{Data Acquisition}

All scans were acquired on a 3T Siemens TIM Trio scanner with a 12-channel phased-array head coil at the University of California, Santa Barbara. Each data acquisition session included both a diffusion tensor imaging (DTI) scan as well as a high-resolution T1-weighted anatomical scan. The structural scan was conducted with an echo planar diffusion weighted technique acquired with iPAT using an acceleration factor of 2. The diffusion scan was 30-directional with a $b$-value of 1000s/mm$^{2}$ and TE/TR = 94/8400 ms, in addition to two $b0$ images. Matrix size was 128$\times$128 with a slice number of 60. Field of view was 230$\times$230mm$^2$ and slice thickness was 2mm. Acquisition time per DTI scan was 9:09min. The anatomical scan was a high-resolution three-dimensional T1-weighted sagittal whole-brain image using a magnetization prepared rapid acquisition gradient-echo (MPRAGE) sequence. It was acquired with TR = 2300 ms; TE=2.98 ms; flip angle = 9 degrees; 160 slices; 1.10mm thickness.

\subsubsection*{DTI Preprocessing}

DTI is both highly sensitive to subject movement \citep{Yendiki:2013ez} and susceptible to directional eddy currents, which can cause distortions in the brain volume \citep{Jezzard:1998tr}. To address these issues, we performed the following data preprocessing using the FMRIB Software Library (FSL v5.0.8) \citep{Smith:2004eq,Jenkinson:2012dj}. First, individual subject masks of the brain were created with the Brain Extraction Tool (BET) \citep{Smith:2002ef} for use in later registration and correction tools, which require an accurate estimation of the spatial extent of the brain. We applied the EDDY correction tool \citep{Andersson:2016de} which simultaneously models both motion effects and eddy current distortions, and corrects them relative to a $b=0$ image collected at the beginning of the scan.

Next, subject scans were transformed into a common space to compare regional connectivity between subjects. Using FMRIB's Linear Image Registration Tool (FLIRT) \citep{Jenkinson:2001cd,Jenkinson:2002hj}, scans were registered to the anatomical T1 image, and then the anatomical scan was in turn registered to the Montreal Neurological Institute (MNI) space MNI152 template using FMRIB's Nonlinear Image Registration Tool (FNIRT). Motion correction also impacts the effective $b$-matrix directions since the rotated images are no longer aligned with the scanner; therefore, we used the output of EDDY to rotate the $b$-vectors to match the changes induced by the motion correction procedure \citep{Leemans:2009jn}.

Using DSI-Studio (http://dsi-studio.labsolver.org), orientation density functions (ODFs) within each voxel were reconstructed from the corrected scans in native diffusion space in order to minimize sampling distortions \citep{Cieslak2014}. We then used the reconstructed ODFs to perform a whole-brain deterministic tractography using DSI-Studio \citep{Yeh:2013fa}. We generated 1,000,000 streamlines per subject, with a maximum turning angle of 35 degrees\citep{Bassett2011structure} and a maximum length of 500mm\citep{Cieslak2014}. By holding the number of streamlines between participants constant, we use the number of streamlines that connect brain region pairs as an estimate of the strength of the connection and examine individual variability in structural connectivity\citep{Griffa:2013gn}.

\textbf{Table 2}

\subsubsection*{Network Construction}

To examine the relationship between structural connectivity and individual differences in learning rate, we constructed networks for each subject where nodes are atlas regions and edges are the measured connection strength between region pairs \citep{Hagmann:2008gd}.

The nodes of the network were derived from spatially-defined regions of a brain atlas, and we utilized two complementary atlas parcellations to confirm that our results are not specific to the particular regional boundaries chosen by one atlas. First, we used the Harvard-Oxford atlas to allow for direct comparison to functional network studies of this same task \citep{Bassett:2013hs,Bassett2014,Bassett2015}, and we combined the Harvard-Oxford cortical and subcortical atlases into a single 111-region atlas by giving cortical labels precedence whenever a single voxel was assigned to both a cortical and subcortical region. In our intra- vs interhemisphere analysis, we exclude the brainstem region from this atlas since it crosses the midline. As a complementary parcellation, we chose the anatomically-defined AAL atlas, originally developed in Statistical Parametric Mapping (SPM) \citep{TzourioMazoyer:2002bi}, which divides each brain hemisphere into 45 regions. For both atlases, we used a version in MNI-space that was then warped into subject-specific native space using FNIRT. Across both atlases, the edges of the network were derived from streamlines that started and ended between the region pair and excluded streamlines that passed through one or both of the regions.

Weighted connectivity matrices were then generated from the atlases and DTI reconstructions such that the matrix $\mathbf{W}$ contained elements $W_{ij}$ whose values were equal to the number of streamlines with end-to-end connectivity between regions $i$ and $j$.  All diagonal elements in the matrix were set to 0 to eliminate self-connections. To correct for differing region sizes, each matrix element was divided by the sum of the volumes of regions $i$ and $j$ \citep{Hagmann:2008gd}. That is, $B_{ij} = \frac{W_{ij}}{v_{is} + v_{js}}$ where $v_{is}$ is the number of voxels in region $i$ for subject $s$. The resultant connectivity matrix for each subject and scan was then normalized to give a \textit{connection strength} $\mathbf{A}$ such that $A_{ij} = \frac{B_{ij}}{\sum_{i,j}B_{ij}}$, ensuring that all scans had identical total connection strength.

\subsubsection*{Network Statistics}

Based on the functional analysis of this dataset \citep{Bassett2015}, we examined whether individual variability in structural connectivity among distributed regions of the motor and visual systems was correlated with learning rate \citep{Mattar2015}. For both of these systems, we calculated the mean connection strength within the system by averaging the weights of all edges connecting pairs of nodes within the system (see Table~\ref{Tab1}). We report our results both at the single-scan level as well as an average over the 4 scans of each subject. Results are consistent in the two cases.

To analyze the impact of indirect connectivity between motor and visual regions, we computed \emph{walk strength}, a measure of the connection strength between two regions that accounts for indirect paths of varying walk lengths. Here, a walk is defined as a path from one point in the graph to another that may pass along the same edge more than once (Fig.~\ref{fig6}A). Given a graph $\mathbf{G}$ and its adjacency matrix $\mathbf{A}$, $A^n$ provides the connection strength between all pairs of nodes when examining walks of length $n$ \citep{Estrada:2007eo}. For instance, streamlines directly connecting primary visual cortex to primary motor cortex would be a walk of length 1, whereas the combination of streamlines connecting primary motor cortex first to thalamus and then to primary visual cortex would be a walk of length 2. Note that the term ``length'' is used in a topological sense, where walks with more steps are considered to have longer length \citep{Crofts2009}. We base our analysis on a similar metric, \emph{communicability}, which is defined such that walks of all lengths contribute to network communication, but longer walks increasingly contribute less. For an unweighted graph, the network communicability is simply given as $\sum_{n=1}^{\infty}\frac{A^n}{n!}$ \citep{Estrada:2007eo}. In a weighted graph, an additional normalization is needed to prevent highly connected nodes from unduly dominating the estimate \citep{Crofts2009}. A typical solution is to divide all weights $A_{ij}$ by $\sqrt{d_id_j}$ where $d_i$ is the degree of node $i$, given by $d_i=\sum_{k=1}^\infty A_{ik}$\citep{Higham:2007bh}. While communicability provides a single metric of communication between nodes, it does not provide information on the contributions of specific walk lengths. To address this limitation, we define the \emph{walk strength} as the normalized strength of walks of length $n$, which is given as $S^n$ where $S=D^{-\frac{1}{2}}AD^{-\frac{1}{2}}$ and $D=diag(d_i)$, or the matrix whose diagonal is given by the values $d_i$.\citep{Crofts:2008wr}.

\subsubsection*{Statistical Testing}

Analysis was performed in Python using a collection of freely available packages: Numpy/Scipy, Pandas, stastmodels and Jupyter. Correlations reported throughout the paper are Pearson correlations at an $\alpha$ level of $0.05$. Data was corrected for multiple comparisons using Bonferroni, False Discovery Rate \citep{Benjamini:1995ws}, and the form $p < \frac{0.05}{n}$, where $n$ is the number of comparisons.

\section*{Results:}

\subsection*{Visual Streamline Connectivity Correlates with Learning}

\textbf{Figure 3}

\begin{figure}[!ht]
\centering
\includegraphics{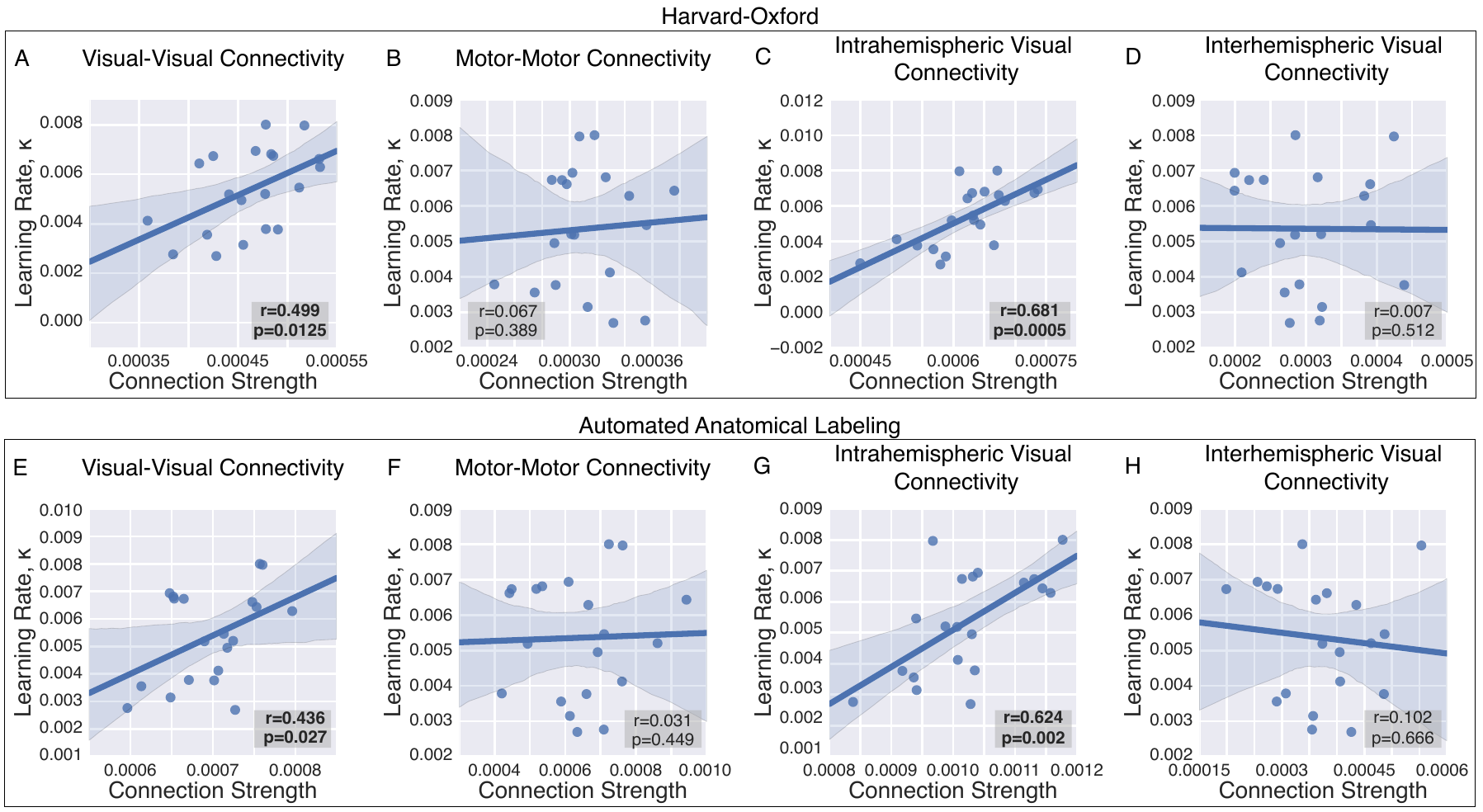}
\caption{\textbf{Correlations between mean connection strength and learning rate across all scanning sessions.}  \emph{(A)} We observe a significant correlation between the average strength of connections linking visual regionss and the learning rate. \emph{(B)} No such relationship is observed for connections linking motor regions. The correlation between learning rate and visual-visual connectivity is largely driven by intrahemispheric \emph{(C)} rather than interhemispheric \emph{(D)} connections. \emph{(E-H)}: We observe the same relationships in the AAL atlas as in the Harvard Oxford atlas for each subset of connections.
\label{mean_scans}}
\end{figure}

Our general aim was to uncover the structural network correlates of individual differences in learning rate for a common visuo-motor task \citep{Wymbs:2015ds}. Because direct connections between motor and visual cortices are not present at this large scale, we separately consider connectivity within motor areas and within visual areas previously identified in a functional analysis of this dataset \citep{Mattar2015,Bassett2015} (see Table~\ref{Tab1} and Fig.~\ref{mv_dti}A--B). We explicitly test the fundamental hypothesis that individuals with greater mean structural connectivity in motor and visual cortices would show faster learning rates ($\kappa$; see Methods) than individuals with less connectivity. We observed a highly significant correlation between visual-visual streamlines and learning rate across all subjects (Pearson correlation coefficient $r=0.50$, with corresponding one-tailed $p$-value of $p=0.0125$, significant after Bonferroni correction; Fig.~\ref{mean_scans}A). In contrast, we observed no significant correlation between motor-motor streamlines and learning rate ($r=0.07, p=0.389$; Fig.~\ref{mean_scans}B).

Within the subset of connections linking visual regions with one another, we expected that connection strength within a given hemisphere would be particularly relevant given that interhemispheric transfer of information is not as relevant in this task as it is in other tasks manipulating perceptual reference frames \citep{Bernier2010} or narrow visual fields \citep{doron2012dynamic}. Consistent with our hypothesis, we found that the observed correlation between the visual-to-visual connection strength and learning rate was largely driven by intrahemispheric streamlines (Pearson correlation coefficient $r=0.68$, $p=0.0005$; Fig.~\ref{mean_scans}C), while no significant correlation was observed among interhemispheric connections ($r=0.01$, $p=0.512$; Fig.~\ref{mean_scans}D).

We verified these same relationships in the AAL atlas (Fig.~\ref{mean_scans}E--H). The structural connection strength among visual-visual region pairs accounts for individual variability in learning rate. It is again more pronounced in both overall visual ($r=0.44$, $p=0.027$) and intrahemispheric visual-visual connectivity ($r=0.62$, $p=0.0002$), while no significant correlation is observed either in motor-motor connectivity ($r=0.03$, $p=0.449$) or in interhemispheric visual connections ($r=0.10$, $p=0.666$).

\subsection*{Reliability of Connectivity-Based Predictors of Learning:}

\textbf{Figure 4}

\begin{figure}[!ht]
\centering
\includegraphics{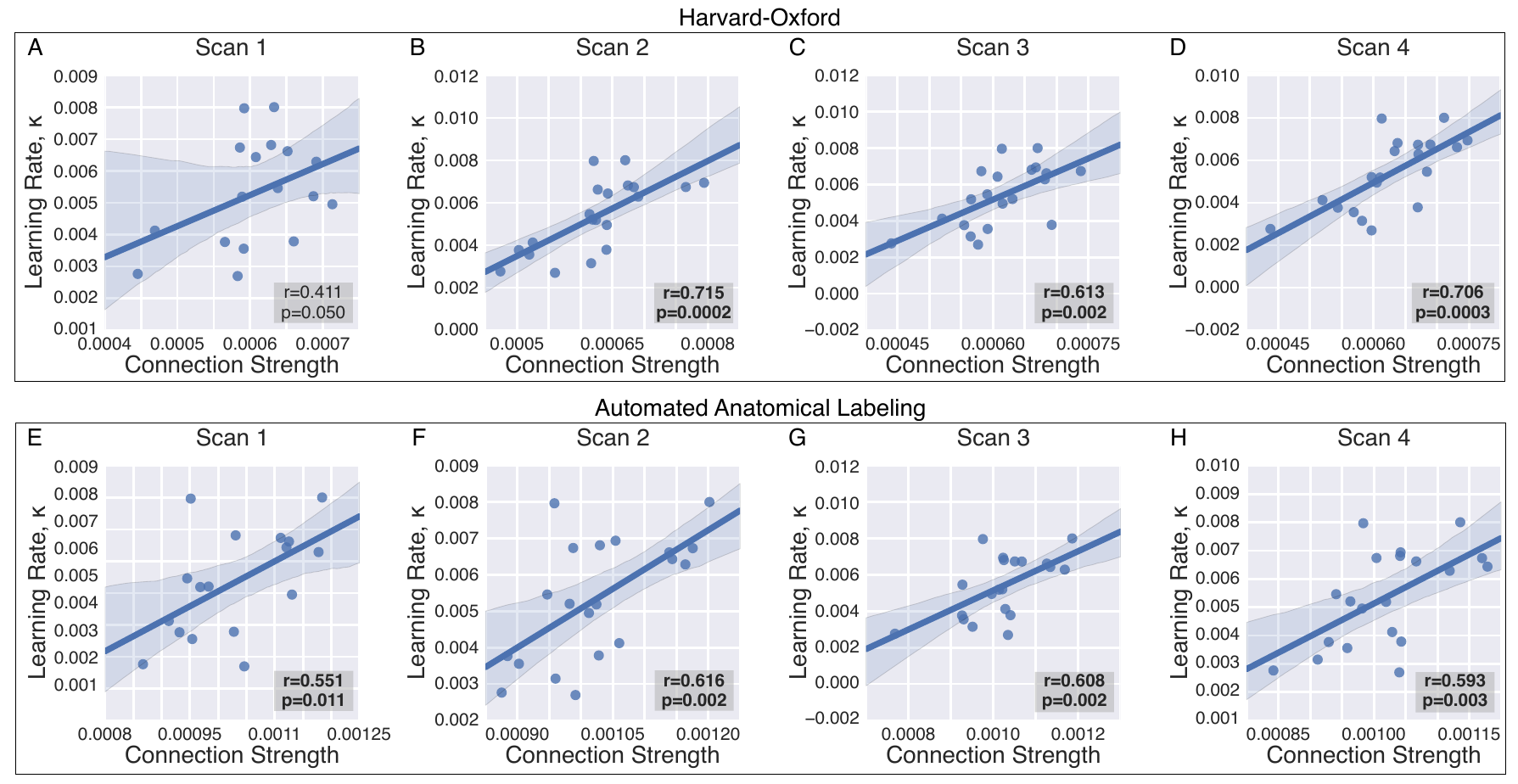}
\caption{\textbf{Connectivity-learning relationship by scan} \emph{(A--D)}: The structural connection strength between intrahemispheric visual-visual region pairs accounts for individual variability in learning rate, and this relationship is stable across the four scanning sessions. This relationship is significant after Bonferroni correction for Scans 2, 3, and 4 in the Harvard-Oxford atlas, with near significance in Scan 1. \emph{(E-H)}: We replicated these results within the AAL atlas, showing significance in all four scan sessions.
\label{individual_scans}}
\end{figure}

Next, we asked whether individual differences in the white matter connections that predicted learning rate would remain constant \citep{LeBihan:2012cy} or change appreciably \citep{Scholz2009,blumenfeld2011diffusion,taubert2012learning} over 6 weeks of practice. We performed the same analysis as before but individually applied to each scan, restricting ourselves to the set of visual intrahemispheric connections (Fig.~\ref{individual_scans}A--D). Across the four scan sessions, we observed a positive relationship between the visual-to-visual connection strength and learning rate: the $p$-values for scans 2, 3, and 4 all pass a Bonferroni correction for $n=4$ tests and scan 1 was close to significant at $p=0.05$. Pearson correlation coefficients and corresponding $p$-values for scan 1 were $r=0.41, p=0.050$, for scan 2 were $r=0.72, p =0.0002$, for scan 3 were $r=0.61, p=0.002$ , and for scan 4 were $r=0.71, p=0.0003$. These results suggest that the connectivity-learning relationship remained constant over 6 weeks of practice.

In addition to being robust across scanning sessions, the connectivity-learning relationship is also robustly observed when we segregated the brain into 90 (rather than 111) regions using a separate atlas. Specifically, using the automated anatomical labeling (AAL) atlas, we observed a significant correlation between learning rate and intrahemispheric visual connection strength across all four scan sessions after Bonferroni correction for $n=4$ tests (Fig.~\ref{individual_scans}E--H). Pearson correlation coefficients and corresponding $p$-values for scan 1 were $r=0.51$, $p=0.011$, for scan 2 were $r=0.62$, $p=0.002$, for scan 3 were $r=0.61$, $p=0.002$, and for scan 4 were $r=0.54$, $p=0.003$.

The scan-independent relationship between learning rate and visual-to-visual connectivity suggests the possibility that visual-to-visual connectivity itself is consistent across the 6 weeks of training, consistent with previous reports in other learning contexts \citep{LeBihan:2012cy}. To directly assess the reliability of visual-to-visual connectivity, we performed two separate analyses: one at the level of white matter streamlines and the second at the level of fractional anisotropy across voxels. First, we computed the intraclass correlation coefficient (ICC)\citep{Shrout:1979kz} to assess the reliability of visual connectivity across scanning sessions across the subset of subjects present for all four scans ($n=17$). Using a two-way ANOVA on visual-to-visual connection strength, we found no main effect of scanning session ($F(3,48)=1.35$, $p=0.27$). Furthermore, the ICC is extremely high ($\text{ICC(1,1)}=0.83$), which indicates the high reliability of visual-to-visual connection strength across scanning sessions. Additionally, we performed voxel level univariate analyses to test for reliability of fractional anisotropy across the whole brain over the 6 weeks of learning. A repeated measures ANOVA was calculated across the four DTI scan sessions. An $f$-omnibus test demonstrated no significant effects ($p>0.05$, FDR corrected). In addition, a paired $t$-test between scans 1 and 4 was performed. There were no significant differences of FA values ($p>0.05$, FDR corrected). These results support the conclusion that white matter microstructure remains consistent over the 4 scans, supporting the observed inter-scan reliability of our results.

\subsection*{Anatomical Specificity of Connectivity-Learning Relationship}

\textbf{Figure 5}

\begin{figure}[!ht]
\centering
\includegraphics{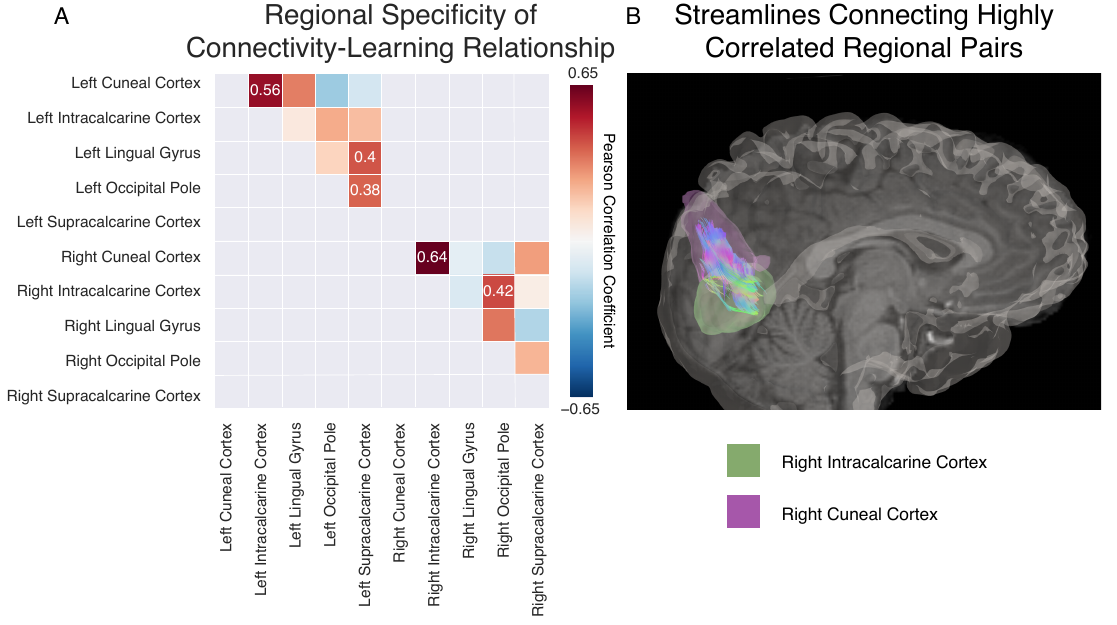}
\caption{\textbf{Anatomical specificity of visual-to-visual connection strength correlation with learning rate.} \emph{(A)}  Significant correlation coefficients (uncorrected) between connection strength and learning rate $\kappa$ for intrahemispheric visual regions are shown in colored boxes. We note relationships that pass a correction for multiple comparisons of the form $p < \frac{0.05}{n}$, where $n$ is the number of comparisons. Gray boxes were not included in the analysis, representing either duplicate entries or interhemisphere connections. \emph{(B)} The reconstructed streamlines are shown for the only region pair that survives FDR correction: the connection between right intracalcarine cortex (green region) and right cuneal cortex (purple region).
\label{regional}}
\end{figure}

To better understand the relationship between intrahemispheric visual connectivity and variability in learning rate $\kappa$, we examined which visual region pairs were driving this effect. This examination had the added benefit of assessing whether different connections predicted behavior differently: although a positive trend was expected given the results in Fig.~\ref{mean_scans}, it is possible that a few smaller regions might show the opposite relationship. To address these questions, for each visual region pair we calculated the Pearson correlation coefficient between the subject learning rate and the mean connection strength between the two regions across scans (see Fig.~\ref{regional}A). We found significant correlations (uncorrected) between learning rate and individual differences in the connections between five pairs of visual regions: right intracalcarine and right cuneal cortex $(r=0.64, p=0.0012)$, right cuneal cortex and right occipital pole $(r=0.42, p=0.032)$, left intracalcarine and left cuneal cortex $(r=0.56, p=0.005)$, left supracalcarine and left occipital cortex $(r=0.38, p=0.049)$, and left supracalcarine and left lingual gyrus $(r=0.4, p=0.039)$. Only the first of these relationships passed FDR correction for multiple comparisons (Fig.~\ref{regional}B).

\subsection*{Role of Indirect Connectivity in Learning Prediction}

\begin{figure}[!ht]
\centering
\includegraphics{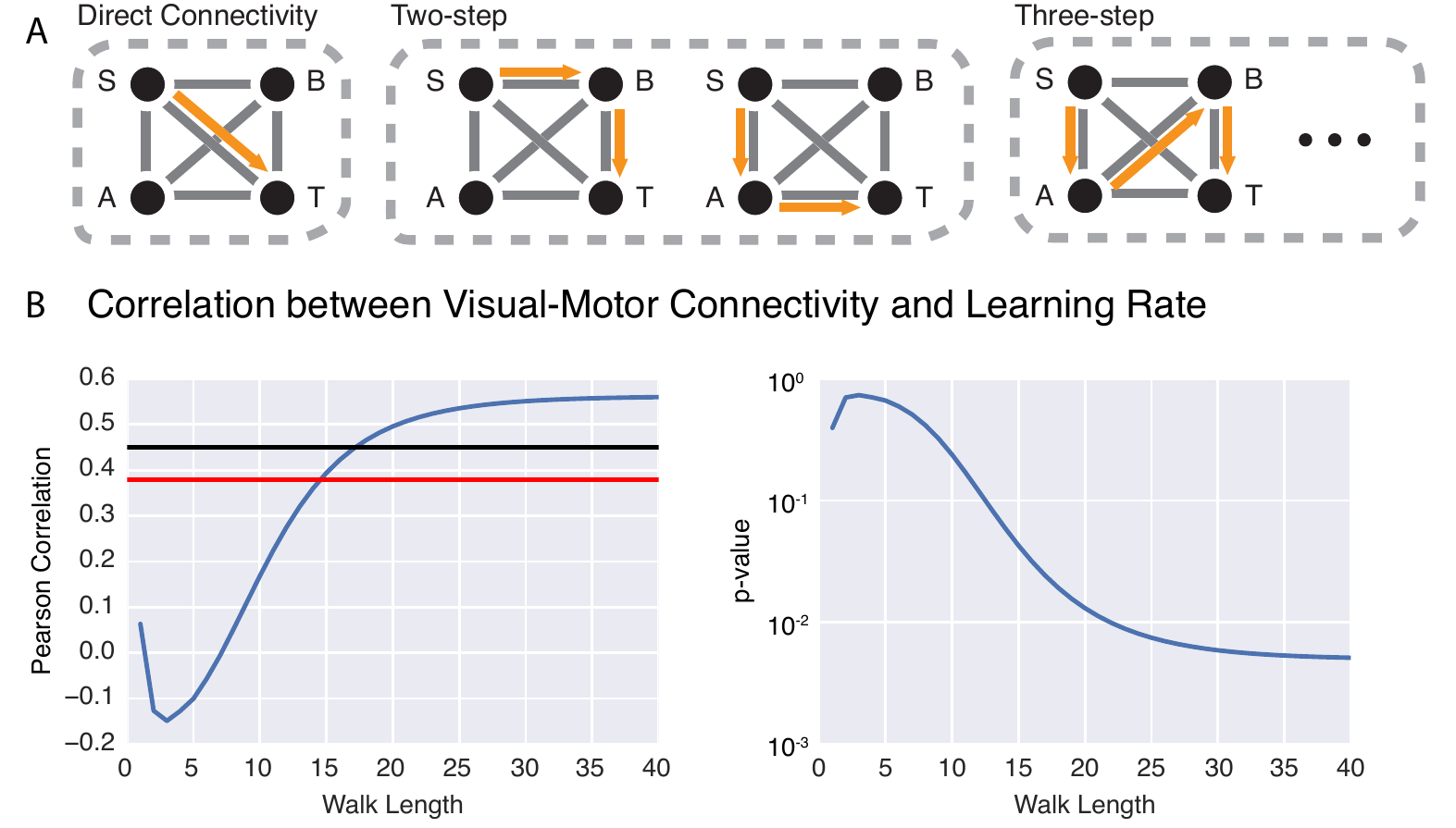}
\caption{\textbf{Indirect connections between visual and motor cortex facilitate learning.} \emph{(A)} Indirect connections between regions of interest can be quantified by walks on a structural network. Consider a toy graph in which paths exist from the source ($S$) to the target ($T$). The eventual flow of information between $S$ and $T$ will not only be influenced by direct connections, but also by indirect walks of length greater than one. \emph{(B)} Correlation between individual differences in motor-visual connection strength at increasing walk lengths (measured by $(D^{-\frac{1}{2}}AD^{-\frac{1}{2}})^n$) and individual differences in learning rate $\kappa$. The correlation becomes significant at a walk length of $n=15$. The red line indicates the $p=0.05$ significance level calculated from the expectation of Pearson correlation coefficients in normal data; the black line indicates the $p=0.05$ significance level calculated from a non-parametric permutation based null model in which node labels have been shuffled uniformly at random.
\label{fig6}}
\end{figure}

Our results have revealed structural correlates in direct connections \emph{within} visual and motor regions; however, prior fMRI studies have linked changes in learning rate to functional connectivity \emph{between} motor and visual areas \citep{Mattar2015,Bassett2015}. To examine structural predictors between regions, we turn to recently developed mathematical techniques in the domain of network science that allow us to directly examine the effects of indirect connectivity (an estimate of polysynaptic transmission potential across extended physical distances) in brain networks. Specifically, we compute variable walk lengths between any two nodes in a network. Direct connections are a walk length of 1, while connections that pass through one intermediary region have a walk length of 2; connections that pass through two intermediary regions have a walk length of 3, and so on (Fig.~\ref{fig6}A). We hypothesized that as we examined sufficiently long walk lengths, the connectivity between motor and visual regions would become increasingly correlated with individual differences in learning rate.

\textbf{Figure 6}

Our results confirm this hypothesis, demonstrating that the length-specific connectivity between motor and visual regions was increasingly correlated with individual differences in learning rate as walk length increased. As shown in Fig.~\ref{fig6}, at walks of length 15, individual differences in walk strength between motor and visual regions were significantly correlated with individual differences in learning rate (Pearson correlation coefficient $r=0.39$, $p=0.004$). As walk length continued to increase, the correlation approached an asymptote which can be observed at $n=40$ with $r=0.56$, $p=0.005$. We confirmed these assessments of statistical significance using a non-parametric null model wherein we shuffled node assignments to ``visual'' or ``motor'' sets, thereby choosing a random set of pseudo visual-motor edges. We examined the correlation at walks of $n=40$ on repeated null model samples, and constructed a 95\% threshold for the correlation coefficient from the null distribution.  We observed that walks of length $n=18$ and beyond all exceeded this threshold, and at $n=40$ our data was significant compared to the null model at $p=0.008$. These results indicate the importance of indirect connections between motor and visual cortices in facilitating the learning of a visuo-motor task.

\section*{Discussion}

In this study, we assess whether individual differences in structural connectivity can account for individual differences in learning a visuo-motor task. Participants practiced a set of ten-element sequences over a six-week period, and we collected structural imaging data during four MRI scanning sessions spaced two weeks apart. We mapped structural connectivity between brain regions in large networks of interest in motor and visual systems, identified by prior assessments of functional neuroimaging data during task performance \citep{Bassett2015}. We observed a significant correlation between visual (but not motor) structural connectivity and learning rate across participants, and this relationship was consistent across the 4 scanning sessions. Interestingly, this correlation was strongest in direct connections among visual regions within the same hemisphere. However, an assessment of network walk strength also revealed a significant correlation between the strength of indirect connections \emph{between} motor and visual cortices and individual differences in learning rate, suggesting the potential importance of physically extended polysynaptic information transmission for skill acquisition.

\textbf{The Relationship Between White Matter Microstructure and Human Behavior.} Our primary hypothesis posited that individual variability in white matter microstructure connecting task-relevant regions would account for individual differences in skill acquisition. Consistent with our hypothesis, we observed a significant relationship between intrahemispheric connections among visual region pairs and variability in learning rate on a discrete sequence production task practiced over the course of 6 weeks. Previous studies have offered preliminary evidence to suggest that structural differences in specific brain regions (although not networks) correlate with individual differences in skill learning \citep{Tomassini2011,Tuch:2005cf}. Our work extends these previous studies by demonstrating that the degree of connectivity within visual regions is correlated with individual differences in learning rate on a simple motor-visual task. While previous studies have focused on regional or tract-specific changes in fractional anisotropy (FA) in white matter, we demonstrate that tractography-based approaches capture individual differences in white matter that directly support skill acquisition. Of note, we do not observe longitudinal changes of fractional anisotropy in our study population over the course of training, suggesting that our diffusion measures of connectivity are remarkably stationary. Not surprisingly then, we found a remarkably consistent relationship between individual differences of connectivity and learning rate across all four DTI scanning sessions. While both tractography and FA-based approaches can reveal important structural differences, a tractography-based approach allows us to leverage network-based tools to understand brain- and system-wide dynamics.

Although we expected structural variability in both visual and motor systems would correlate with individual variability in learning rate, we only found a significant relationship with intrahemispheric connections among visual regions. We speculate that this may be due to the nature of the motor task itself. The participants learned to quickly press one of five buttons following a visual cue. This specific action (a button press) is not a particularly novel movement for these participants, all of whom have already developed a wide variety of dexterous skills such as typing. Due to the ubiquity of this action over the course of development, the structural connectivity within motor cortex may already be at a ceiling, obscuring any correlation with learning. Alternatively, it is possible that changes in motor connectivity with learning may only be measured at smaller spatial scales. In contrast to the simple button press, the more challenging skill that the subjects mastered was the spatial mapping between visual stimuli and motor commands. It is intuitively plausible that the ability to learn this mapping efficiently is fundamentally dependent on visual resources for detailed encoding of spatial information.  Indeed, a wide range of visuo-motor tasks have demonstrated strong reliance on occipital areas in mapping arbitrary stimuli with specific motor responses as well as sequences of responses \citep{Grafton1994,Grafton1995,Diedrichsen2015,Wiestler2014}.

\textbf{Structure is Consistent Across Scanning Sessions}. Our results demonstrated a relationship between individual variability in learning rate and connection strength between visual regions that was consistent across the four scanning sessions. This consistency is particularly interesting in light of prior work showing changes in brain network connectivity as a function of experience-dependent plasticity \citep{Lindenberger:2006kx,PascualLeone:2005hl}. Indeed, researchers actively debate the time scales at which these structural changes occur \citep{May:2011jw,Holtmaat:2009hp,Keller:2015cz} and whether these changes can be detected using current diffusion weighted imaging techniques \citep{Thomas:2013ft,Lovden:2013dx}. Some of the most well-known experience-dependent plasticity changes have been reported from motor learning tasks \citep{Zatorre:2012en}. Using multi-week training paradigms in juggling, some of these studies have identified both volumetric changes in visual and parietal cortices \citep{Draganski:2004hs,Scholz2009} and fractional anisotropy changes in the posterior intraparietal sulcus \citep{Scholz2009}. Complementary work has examined structural correlates for professional piano players, identifying volumetric differences in motor and parietal regions \citep{Gaser:2003vm} as well as structural connectivity differences in DTI data within the corticospinal tracts that connect motor cortex with the brainstem and spinal cord \citep{Bengtsson:2005fq}. These training induced changes may arise from activity-dependent myelination \citep{Fields2015}, which in turn may contribute to the observed changes in functional connectivity during long-term motor learning \citep{Sampaio2015}. However, unlike juggling or extensive piano practice, our participants did not train on a complex visuo-motor task, but instead, they learned a pairing between a visual cue and a required finger movement for a set of six sequences. In the context of this fine-motor training, we observed a stable relationship between visual connectivity and subject learning rate across all four scans, independent of the number of trials practiced.

\textbf{A Putative Role for Physically Extended Polysynaptic Connections}.
Because prior work in functional neuroimaging has linked changes in learning rate to functional connectivity \emph{between} motor and visual areas \citep{Mattar2015,Bassett2015}, we directly assessed indirect connectivity defined as a variant of network communicability that we called \emph{walk strength}. This metric computes variable walk lengths where paths between two nodes can have increasing numbers of intermediary steps. For example, a two-step walk could be taken from visual cortex through thalamus to motor cortex. We found that as walk length increased, individual differences in motor-visual connectivity were increasingly correlated with learning rate. These results suggest a role for physically extended sets of polysynaptic connections between motor and visual cortices that support the acquisition of this visuo-motor skill. Such a role is consistent with previous work in computational neuroscience highlighting the role of highly structured circuits in sequence generation and memory \citep{Rajan2016,Hermundstad2011}. Indeed, in computational models at the neuron level, architectures reminiscent of chains \citep{Levy2001,Fiete2010} and rings are particularly conducive to the generation of sequences. Our results complement these insights at small spatial scales to suggest that long-distance (chain-like) paths at the large scale of white matter tractography are supportive of sequence production. In future, it may be interesting to assess the generalizability of these results across other sequential learning tasks, and to determine the degree to which additional measurements of indirect connectivity \citep{Goni2014} may differentially relate to learning rate, performance accuracy, and reaction time \citep{Tuch:2005cf}.

\subsection*{Methodological Considerations}
First, it is important to note that in this study, we rely on DTI and white matter tractography to estimate subject-specific and whole-brain structural connectivity. However, it is important to note that DTI-based tractographic reconstructions present a number of limitations. Among these is the tendency of current methods to present false positives and false negatives when compared to histological studies \citep{Thomas:2014ig,Reveley:2015kt}. However, diffusion imaging remains the only reliable method for studying human white matter structure noninvasively. Moreover, we expect potential tractography biases to be consistent across subjects, allowing us to accurately access individual differences in white matter architecture and its relationship to behavior. Second, it is also important to note that it is not possible using these techniques to decipher the number of synapses present along the tracts between two regions, nor is it possible to decipher the number of synapses present along long-distance paths in the network. Thus, while the data supports a role for physically extended polysynaptic pathways, it does not directly speak to their microstructure. Third, it is important to note that while we hypothesized that interhemispheric connections would be less important for this task than for other tasks that require the manipulation of perceptual reference frames \citep{Bernier2010} or that utilized a single visual hemifield \citep{doron2012dynamic}, it is nevertheless possible that interhemispheric connections also play a role. It will be important in the future to implement higher resolution diffusion imaging to clarify the potential role of interhemispheric connections in the learning of this novel visuo-motor skill. Fourth, it is interesting to ask whether the structural drivers of individual differences in learning rate are anatomically co-located with observed changes in functional connectivity during task performance. In fact, evidence suggests that this is not the case, and that instead regions that show individual differences in structural connectivity that are predictive of individual differences in learning rate are not the same as the regions that display changes in functional connectivity with training \citep{Bassett2015}. Together, these data suggest that further study is needed to understand the relationships between individual differences in structural connectivity and functional connectivity, and how they relate to gross changes in behavior or to individual differences in learning rate.  Finally, we note that the lack of longitudinal changes in the strength of connectivity (measured both with fractional anisotropy and with the number of reconstructed streamlines between pairs of large-scale brain regions) could be explained either by neuroscientific or methodological factors. It is important to note that with this particular data set, we are unable to determine the origin of this consistency with complete confidence.

\subsection*{Conclusion} We identified variability in structural connectivity that accounts for individual differences in learning rate over six weeks of training on a visuo-motor skill. Our analysis revealed direct connections among intrahemispheric visual regions as well as indirect connections between visual and motor cortices that suggests an underlying mechanism for differences in behavior. Clinically, these results offer novel biomarkers that may prove useful in predicting the time scales of motor rehabilitation following stroke and brain injury. In particular, because individuals with greater visual connectivity show swifter learning rates, a clinician may be able to predict the rate at which a patient will re-learn a motor skill after a stroke based on the degree to which their visual system (and its indirect connections with the motor system) remain intact. More generally, our results may inform personalized training paradigms for healthy individuals; individuals with greater visual connectivity -- and greater strength of indirect connectivity between motor and visual systems -- may require less training to obtain the same proficiency as an individual with lesser connectivity and greater training. While speculative at this point, these possibilities motivate future work in clarifying the utility of white matter architecture in optimizing visuo-motor training across healthy and injured populations.

\section*{Acknowledgments} D.S.B. and A.K. acknowledge support from the John D. and Catherine T. MacArthur Foundation, the Alfred P. Sloan Foundation, the Army Research Laboratory and the Army Research Office through contract numbers W911NF-10-2-0022 and W911NF-14-1-0679, the National Institute of Mental Health (2-R01-DC-009209-11), the National Institute of Child Health and Human Development (1R01HD086888-01), the Office of Naval Research, and the National Science Foundation (BCS-1441502, BCS-1430087, and PHY-1554488). S.T.G. and N.F.W. acknowledge support from the National Institute of Neurologic Disease and Stroke (1-P01-NS44393). The content is solely the responsibility of the authors and does not necessarily represent the official views of any of the funding agencies.

\bibliography{bibfile_v4}
\bibliographystyle{apalike}

\newpage

\begin{table}[ht]
\begin{centering}
\begin{tabular}{p{3cm} l l l l}
\hline
& Scan 1 & Scan 2 & Scan 3 & Scan 4 \\
\hline
MIN sequences & 50 & 110 & 170 & 230 \\
MOD sequences & 50 & 200 & 350 & 500 \\
EXT sequences & 50 & 740 & 1,430 & 2,120 \\
\hline
\end{tabular}
\caption{\label{t:training} Number of trials practiced of each sequence type at the start of each scanning session. }
\end{centering}
\end{table}

\newpage

\begin{table}
\hfill{}
\begin{tabular}{|l|l|}
\hline
\small{Motor} & \small{Visual}\\
\hline
\small{L,R Precentral gyrus}              & \small{L,R Intracalcarine cortex}\\
\small{L,R Postcentral gyrus}             & \small{L,R Cuneus cortex}\\
\small{L,R Superior parietal lobule}      & \small{L,R Lingual gyrus}\\
\small{L,R Supramarginal gyrus, anterior} & \small{L,R Supercalcarine cortex}\\
\small{L,R Supplemental motor area}       & \small{L,R Occipital Pole}\\
\small{L~~ Parietal operculum cortex}       & \\
\small{R~~ Supramarginal gyrus, posterior}  & \\
\hline
\end{tabular}
\hfill{}
\caption{Brain areas in motor and visual systems derived directly from functional neuroimaging studies of the same task \citep{Bassett2015}.}
\label{Tab1}
\end{table}

\newpage

\newpage

\newpage

\newpage

\newpage

\end{document}